\documentclass[twocolumn,showpacs,preprintnumbers,amsmath,amssymb,prb]{revtex4}

\usepackage{epsfig}
\usepackage{graphicx}
\usepackage{dcolumn}
\usepackage{bm}
\usepackage{subfigure}
\usepackage{color}
\newcommand{\nba}[1]{}

\def\oreo6{ReO$_6$}
\def\reo3{ReO$_3$}
\def\nkto{Na$_{0.82}$K$_{0.18}$TaO$_3$}
\def\nto{NaTaO$_3$}
\def\im3{Im$\overline 3$}
\def\pm3m{Pm$\overline 3$m}
\def\p4mbm{P4/mbm}

\begin{document}


\title{Local structure of \reo3~at ambient pressure from neutron total scattering study}

\author{E.~S.~Bo\v{z}in,$^{1}$ T. Chatterji,$^2$ and S.~J.~L.~Billinge$^{1, 3}$}

 \affiliation{$^1$Department of Condensed Matter Physics and Materials Science, Brookhaven National Laboratory, Upton, NY 11973, USA}
 \affiliation{$^2$ Institut Laue-Langevin, 6 rue Jules Horowitz, BP 156
Grenoble cedex 9, France}
 \affiliation{$^3$Department of Applied Physics and Applied Mathematics, Columbia University, New York, NY 10027, USA}

\date{\today}

\begin{abstract}
{
A
hypothesis that the local rotations of ReO$_6$ octahedra persist in the
crystallographically untilted ambient phase of \reo3 is examined by the
high-resolution neutron time-of-flight total scattering based atomic
pair distribution function analysis.  Three candidate models
were tested, \pm3m, \p4mbm, and \im3, for the local structure of
\reo3 at ambient pressure and 12~K,
and both quantitative and qualitative assessment of the data were performed.
No evidence for large local octahedral rotations was found, suggesting that the
local and the average structure are the same (\pm3m) as normally assumed.}
\end{abstract}


\maketitle
%
\section{Introduction}
%
Among d-electron metallic conductors \reo3 has a simple
perovskite-like structure and its conductivity is comparable
to that of a noble metal such as Ag.~\cite{ferre;jpcs65,king;ssc71} Although the
electron-phonon coupling constant is not very small,
\reo3 surprisingly does not show superconductivity down
to 20~mK.~\cite{allen;prb93} \reo3 seems to belong to
the normal class of conventional band Fermi liquids with
electron-phonon interaction dominating the
resistivity.~\cite{allen;prb93}

%
\begin{figure}[tb]
\begin{center}$\,$
\epsfxsize=2.7in
\epsfbox{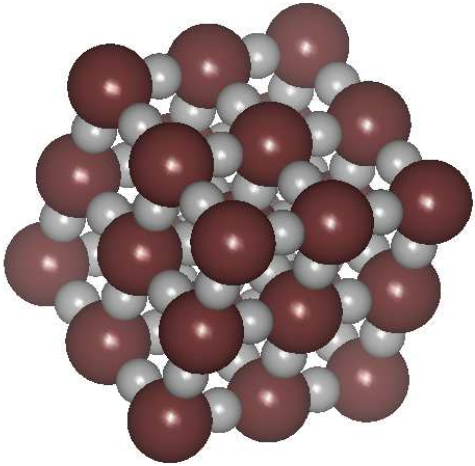}
\end{center}
\caption{Structural motif of \reo3 (space group \pm3m):
Re is shown as large spheres, while small spheres represent O.}
\protect\label{fig;structure}
\end{figure}
%

\reo3 crystallizes in the cubic space group \pm3m with the
undistorted perovskite-like DO$_9$ type structure (ABO$_{3}$,
Fig.~\ref{fig;structure}) comprised of a network of corner-shared
\oreo6 octahedra and with an empty A-site, and is in fact the
simplest material containing BO$_{6}$ octahedra. \reo3 is
very special among the multitude of systems possessing the perovskite-based
structure, in that its untilted cubic \pm3m structure is extremely
stable at ambient pressure and at all temperatures from liquid-helium
temperature~\cite{fujim;ssc80} up to its melting point at
673~K.~\cite{kuzmi;jpcm96,stach;prb97} Electronic band structure
calculations investigated this extraordinary structural
stability~\cite{stach;prb97} and suggested that metallic bonding
plays an important role.~\cite{yu;prb05} The open crystal structure
makes \reo3 suitable for doping or compression under high-pressure.
\reo3 attracted considerable attention recently due to observation of weak negative
thermal expansion (NTE) below room temperature.~\cite{chatt;prb08}
NTE is rarely found in metals, and in \reo3 is rather sensitive to impurity induced disorder.~\cite{rodri;jap09}

A structural phase transition was discovered in \reo3 upon application
of pressure when the pressure induced anomaly in the Fermi
surface was observed in measurements of De Haas-van Alphen
frequencies,~\cite{razav;ssc78} and was characterized in detail
by means of x-ray and neutron
diffraction.~\cite{schir;prl79,axe;prb85,jorge;prb86}
At room temperature \reo3 undergoes a pressure-induced second order
phase transition at $p_{c} = 5.2~kbar$, to an intermediate tetragonal
phase (\p4mbm) over a narrow pressure range, and further to a cubic
(\im3) phase that persists up to relatively high pressures while
the \oreo6 octahedra remain almost undistorted,~\cite{jorge;jac04}
but considerably tilted.~\cite{schir;prb84} The driving force for the
transition is argued to be the softening of the M3 phonon mode involving
a rigid rotation of \oreo6 octahedra. It is a continuous transition with
the rotation angle as an
order parameter.~\cite{jorge;prb86} The pressure-temperature
phase diagram has been recently established,~\cite{chatt;ssc06}
indicating that value of  $p_{c}$ decreases with decreasing
temperature, leveling at around 2.4~kbar at base temperature.
More recently, a study of the pressure induced phase transition
showed that in nanocrystalline \reo3, the sequence of phases is
different and with generally lower $p_c$ values than for bulk
samples.~\cite{biswa;jpcm07}

This canonical view has been challenged by the
reexamination~\cite{house;prb00} of the structural phase transition
in \reo3 using the x-ray absorption fine structure (XAFS) method
sensitive to the nearest neighbor information. This investigation suggests
that even at ambient pressure the Re-O-Re bond angle deviates from 180$^{\circ}$
by about 8$^{\circ}$. In this newly proposed view, \reo3 only appears
cubic in the analysis of the Bragg intensities, while the local structure
is suggested to be heavily distorted. In this picture the phase transition at $p_c$ is then a
tilt order-disorder transition.~\cite{house;prb00}
Similar conclusions were drawn from the study of XAFS spectra of
antiferrodistortive perovskites \nkto and \nto.~\cite{recha;prl94}
The results reported by Houser and Ingalls\cite{house;prb00} are in contrast
to XAFS results reported earlier that suggested agreement between the
local and the average structure,~\cite{house;pb95,dalba;jpcm95,kuzmi;jpcm96}
possibly due to the different approaches to XAFS data
analysis taken in the different studies.~\cite{kuzmi;prb93,house;prb00}
%
\begin{figure}[tb]
\begin{center}$\,$
\epsfxsize=3.2in
\epsfbox{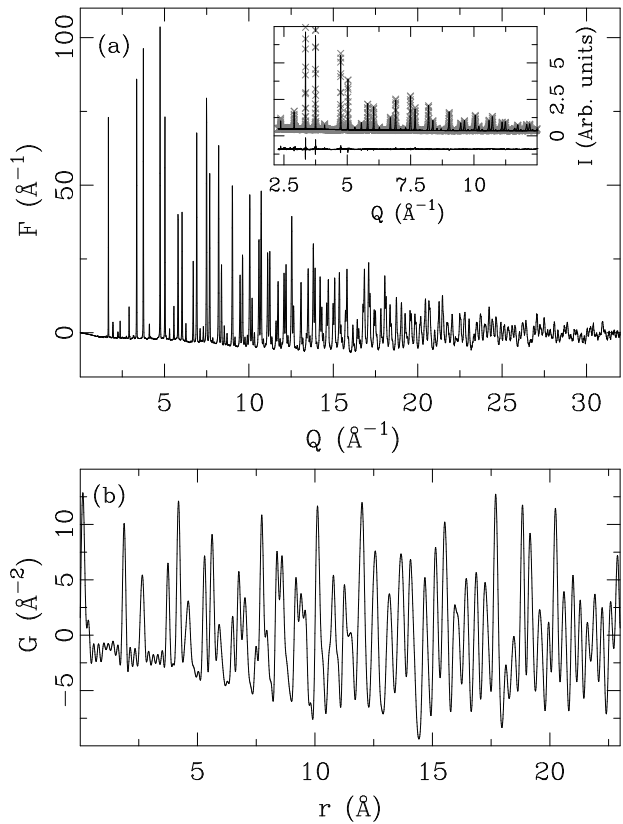}
\end{center}
\caption{Neutron experimental data of \reo3 at 12~K and at ambient
pressure: (a) reduced total scattering structure function, $F(Q)$, and
(b) corresponding atomic PDF, $G(r)$. Inset shows Rietveld refinement
using \pm3m model (solid line) of the normalized intensity (symbols)
with the difference curve offset for clarity.}
\protect\label{fig;experiment}
\end{figure}
%

We have carried out a neutron total scattering study of \reo3 at
ambient pressure and low temperature. Based on the same data
we performed both conventional Rietveld analysis, based on Bragg
intensities, yielding the average crystallographic structure, and the
atomic pair distribution function (PDF) analysis, which is a direct
space method that includes both Bragg and diffuse scattering information,
yielding structural information on local, intermediate and long range
scales.\cite{egami;b;utbp03,billi;jssc08,young;jmc11} This approach allows structure to be assessed on various lengthscales from the same data.
Our analysis rules out the existence of large local tilt amplitudes in \reo3.

%
\section{Experimental}
%
In this study we used 3 grams of commercially available
(Sigma-Aldrich) \reo3 sample in the form of a loose powder.
Neutron time-of-flight  powder diffraction measurements
were carried out using the high-resolution NPDF diffractometer
at the Manuel Lujan Neutron Scattering Center at Los Alamos
National Laboratory. The sample was sealed in a vanadium tube
with He exchange gas, and cooled down to 12~K using a closed
cycle He refrigerator. Raw data were corrected for experimental
effects such as sample absorption and multiple scattering
using the program PDFgetN,~\cite{peter;jac00} to obtain the
total scattering structure function, $S(Q)$. This contains
both Bragg and diffuse scattering and therefore information about
atomic correlations on different length scales. The PDF, $G(r)$,
is obtained by a Fourier transformation according to
$G(r) = {2\over\pi}\int_0^{\infty} Q[S(Q)-1]\sin Qr\>dQ$, where $Q$
is the magnitude of the scattering vector. The PDF gives the
probability of finding an atom at a distance $r$ away from another
atom. The reduced total scattering structure function, $F(Q)=Q[S(Q)-1]$,
of \reo3 at 12~K is shown in Fig.~\ref{fig;experiment}(a) and
the resulting PDF, $G(r)$, in Fig.~\ref{fig;experiment}(b).
PDFs refined in this study were produced using an upper limit of
integration in Fourier transform $Q_{max}$ of 32~\AA$^{-1}$.
%
\begin{table}[tb]
\caption{Summary of structural refinements of \reo3 data.
Atomic displacement parameters, U, are in units of \AA$^{2}$. Derived interatomic distances $d$ and
octahedral rotation angles $\varphi$ are given at the bottom. Numbers in parentheses are ESDs obtained from fitting.}
\label{tab1;structuralresults}
\begin{tabular}{ccccc}
\hline \hline
&Rietveld &\pm3m&\p4mbm&\im3 \\
\hline
a (\AA)&3.7499(2)&3.7508(2)&5.3101(2)&7.5012(3)\\
c (\AA)&-&-&3.7430(3)&-\\
U(Re) &0.0016(3)&0.0011(1)&0.0010(1)&0.0010(1) \\
x(O)&-&-&0.2400(6)& \\
y(O)&-&-&-& 0.2520(7)\\
z(O)&-&-&-& 0.2417(6)\\
U(O)&0.0057(5)&0.0053(3)&0.0031(3)&0.0031(3) \\
U$_{par}$(O)&0.0027(5)&0.0026(3)&-&- \\
U$_{perp}$(O)&0.0073(3)&0.0061(2)&-&- \\
R$_{wp}$ (\%)&2.97&7.89&7.87&8.21\\
\hline
d1$_{Re-O}$ (\AA)&1.8749(1)&1.8754(1)&1.8714(1)&1.8764(2)\\
d2$_{Re-O}$ (\AA)&-&-&1.8789(2)&-\\
d1$_{O-O}$ (\AA)&2.6516(1)&2.6522(2)&2.6520(3)&2.6197(3)\\
d2$_{O-O}$ (\AA)&-&-&2.6572(4)&2.6855(3)\\
$\varphi$ (deg)&0&0&2.3(2)&2.1(2)\\
\hline\hline
\protect\label{tab;results}
\end{tabular}
\end{table}
%
%
\begin{figure}[tb]
\begin{center}$\,$
\epsfxsize=3.2in
\epsfbox{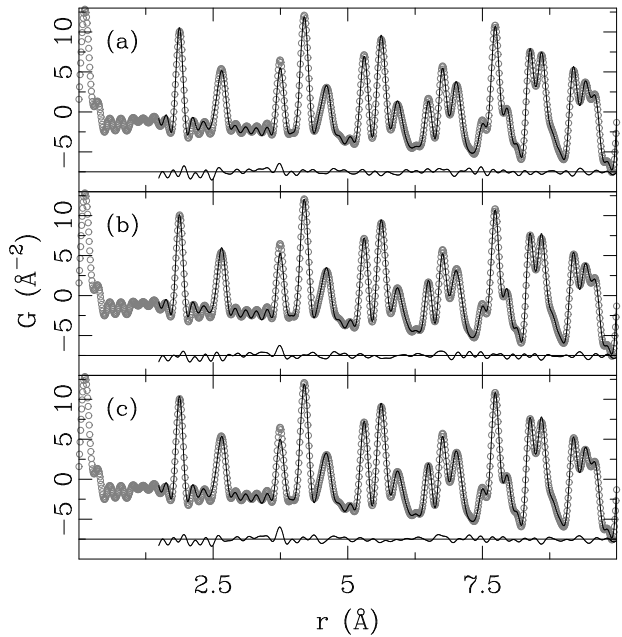}
\end{center}
\caption{Refinement of various models (solid black lines) to the 12~K neutron
PDF data (open symbols):
(a) \pm3m model, (b) \p4mbm model, and (c) \im3 model. The difference
curves are offset for clarity.}
\protect\label{fig;fits}
\end{figure}
%

The average crystal structure was verified by taking the standard Rietveld
refinement approach in reciprocal space using the EXPGUI~\cite{toby;jac01}
platform operating the program GSAS.~\cite{larso;unpub87} The fit
of the \pm3m model is shown as an inset to Fig.~\ref{fig;experiment}(a).
The local structure was studied by refinements of structural models to the
experimental PDF using the program
PDFgui.~\cite{farro;jpcm07} The details of the PDF method are provided
elsewhere.~\cite{egami;b;utbp03}

The Rietveld refinements were carried out on the ambient structure model, \pm3m
space group: Re at 1a (0,0,0) and O at 3d (0.5,0,0). The PDF local structural refinements were
carried out over the range 1.5-10.0~\AA\ for three different candidate
structures: conventional \pm3m, and two models that allow rigid rotations
of \oreo6 octahedra: \p4mbm (Re at 2b (0,0,0.5), O at 2a (0,0,0) and
at 4h (x,x+0.5,0.5)), and \im3 (Re at 8c (0.25,0.25,0.25), O at
24g (0,y,z)).
%
\begin{figure}[ftb]
\begin{center}$\,$
\epsfxsize=3.2in
\epsfbox{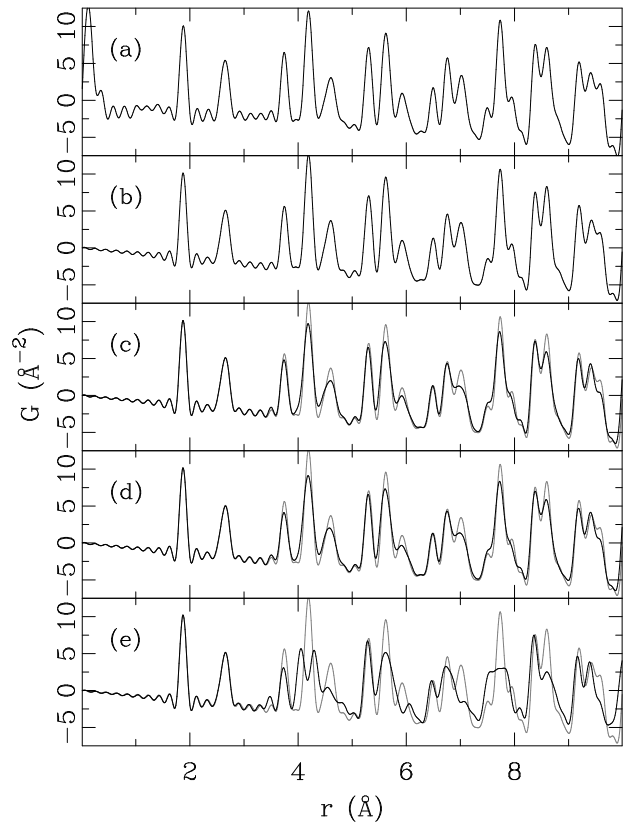}
\end{center}
\caption{Visual comparison of \reo3 PDFs: (a) neutron PDF data,
and various calculated PDFs of competing models: (b) ambient pressure
\pm3m, (c) 5.2~kbar \p4mbm, (d) 5.2~kbar \im3, and (e) 7.3~kbar \im3.
In panels (c)-(e) gray-line PDF profile is that of \pm3m shown in
panel (b), and is given for comparison. See text for details.}
\protect\label{fig;models}
\end{figure}
%

%
\section{Results and Discussion}
%
The results of the Rietveld refinements are shown in the inset
to Fig.~\ref{fig;experiment}(b) and Table~\ref{tab1;structuralresults}.
They reproduce well literature results.~\cite{schir;prl79,axe;prb85,jorge;prb86,rodri;jap09}

We now consider the local structure measured by the PDF.
The fits are shown in Fig.~\ref{fig;fits} (a)-(c),
and the results summarized in Table~\ref{tab;results}.
The PDF does not presume any periodicity and it is therefore possible to
refine lower symmetry structures than the crystallographic model where it
is warranted.  This may be the case when the crystal structure is an 
average over local domains of lower symmetry, as for example in the
ferroelectric BaTiO$_3$.\cite{kwei;ferro95}

As apparent from the figure, all three models provide good
fits. The fact that the crystal structure model provides a good fit with a
good agreement factor suggests that there is no evidence for a lower symmetry local structure.
However, as evident in the table, the good fit benefits from the model being allowed
to refine anisotropic ADPs which are significantly enlarged in directions perpendicular to the
Re-O bond.  It is therefore interesting to consider models that allow for rotations of the octahedral units such as
the symmetry lowered phases observed at high pressure.
The \im3 model, observed at the highest pressures, yields a quantitatively worse fit despite having more refinable
parameters than the crystallographic model. On the other hand, the tetragonal \p4mbm model with isotropic thermal factors imposed gives a comparable agreement to the one obtained using the crystallographic \pm3m model with anisotropic ADPs.
The fits to PDF data are shown in Fig.~\ref{fig;fits}(a-c) for \pm3m, \p4mbm, and \im3 models respectively, and the results are presented in Table~\ref{tab1;structuralresults}.

The PDF measures the instantaneous structure and it is not possible to distinguish explicitly between local tilts that are static and those that are dynamic.   If the tilts are dynamic as we presume here, the tilt angle extracted from the fits of the lower symmetry models yields an average tilt amplitude due to correlated motions of the atoms locally, i.e., rigid rotations of the octahedra, and is a useful quantitative measure of this amplitude.
The octahedral rotation angles, $\varphi$,~\cite{jorge;prb86} obtained from the PDF refinements are $2.3^\circ$ and $2.1^\circ$ for the \p4mbm and \im3 structures, respectively. These are much smaller than $\sim 4.9^\circ$ (equivalent to $\sim 8^\circ$ tilts) reported from XAFS,~\cite{house;prb00} and smaller than values obtained in the pressure stabilized tilted crystal structures.\cite{jorge;prb86}  PDFs calculated with the EXAFS tilt angles gave qualitatively worse fits to the data. The values refined are sufficiently small to suggest that the local tilting is coming from dynamic fluctuations of a low-energy rigid unit tilting mode, as proposed as a mechanism for the observed negative thermal expansion.~\cite{chatt;prb08,wdowi;prb10}

\nba{NOTE: Simon, there should be a bit of discussion here along the lines of rotation angle magnitude at low T, zero point motion, dynamic tilts, and why is \pm3m still the best model. Please note also that in the table for \pm3m model there are anisotropic ADPs, and their isotropic equivalent (refined in separate refinements, obviously. The anisotropic ADPs are enlarged perpendicular to the Re-O bond, suggesting rotations (but that could originate from other factors that we need to mention), and this is then picked up by the tilted models. We should state something about the upper limit for these rotations being set by PDF, and that it is much smaller than what XAFS suggest.}

We have further calculated the PDFs based on data published in the
literature for the three models of interest,~\cite{jorge;prb86} for
comparison keeping the isotropic atomic displacement parameters and the scale
factor the same as those from the PDF \pm3m refinement reported here.
The structural parameters were taken for cases of ambient \pm3m,
5.2~kbar \p4mbm and \im3, as well as for 7.3~kbar \im3 that corresponds
to the stable high-pressure phase case. These calculated PDFs, and
the experimental PDF profile, are shown in Fig.~\ref{fig;models},
and reveal what kind of changes are to be expected in the PDF in the case of \oreo6 octahedral rotations being present locally.
At room temperature, at 5.2~kbar average octahedral rotations of 3.0$^{\circ}$ were
observed, while at 7.3~kbar the rotation angle value rises to
6.6$^{\circ}$.~\cite{jorge;prb86} It is quite apparent from the
calculated PDFs shown in the figure that even for a small rotation
angle there are appreciable changes in the local structure, and
these are rather similar for the two models that allow tilting.
While the first two PDF peaks do not change at all, reflecting the
rigidity of the \oreo6 octahedra,~\cite{dalba;jpcm95} changes start to appear at higher
distances as the \oreo6 network gets distorted. For a larger tilt magnitude,
the changes in the local structure become much more dramatic
Fig.~\ref{fig;models}(e). By visual inspection and comparison of the
features seen in the calculated PDFs, we can immediately rule out the
larger tilt angle case, as there is no resemblance to the observed PDF.

%
\section{conclusion}
%
In summary, we used the PDF method to test a hypothesis from an earlier EXAFS study\cite{house;prb00} 
that the {\it local} rotations
of \oreo6 octahedra persist in the ambient pressure, crystallographically untilted
phase of \reo3. The high-resolution neutron time-of-flight total scattering
based atomic pair distribution function analysis was carried out on a
dataset collected at 12~K under ambient pressure conditions. Three candidate models
were tested, \pm3m, \p4mbm, and \im3, and quantitative and qualitative
assessment of the data were performed. No evidence was found for local octahedral
rotations of a magnitude comparable to those seen in the distorted high
pressure phase and beyond what might be expected from thermal and quantum zero point motion, suggesting that the local and the average structure
are the same (\pm3m) as normally assumed.
The PDF results shown here emphasize again the importance of using multiple
complementary techniques in addressing delicate structural issues at the
nanoscale.\cite{billi;s07}

\begin{acknowledgments}
%
%
ESB acknowledges useful
discussions with John Provis, Efrain Rodriguez, and Anna Llobet.
Work at Brookhaven National Laboratory was supported by the Office
of Science, U.S. Department of Energy (OS-DOE), under contract
no. DE-AC02-98CH10886. This work has benefited from the use of NPDF
at the Lujan Center at Los Alamos Neutron Science Center, funded by DOE Office of Basic
Energy Sciences. Los Alamos National Laboratory is operated by Los
Alamos National Security LLC under DOE contract DE-AC52-06NA25396.
The upgrade of NPDF has been funded by the NSF through grant DMR 00-76488.
%
%
\end{acknowledgments}
%
%


\end{document}